\documentclass{article}

\usepackage[final]{neurips_2019}

\usepackage{microtype}
\usepackage{graphicx}
\usepackage{subfig}
\usepackage{booktabs}
\usepackage{mathtools}
\usepackage{nicefrac}

\usepackage{hyperref}

\usepackage{float}
\usepackage{wrapfig}

\usepackage{amsmath,amssymb}
\DeclareMathOperator{\E}{\mathbb{E}}

\usepackage{pgfplots}
\usepgfplotslibrary{fillbetween}
\usepgfplotslibrary{units}

\definecolor{COLR1}{RGB}{170, 210, 237}
\definecolor{COLR2}{RGB}{106, 190, 198}
\definecolor{COLR3}{RGB}{100, 133, 150}
\definecolor{COLR4}{RGB}{95, 100, 110}
\definecolor{COLR5}{RGB}{88, 65, 70}
\definecolor{COLavg}{RGB}{247, 106, 91}
\definecolor{COLZeroLine}{RGB}{0, 0, 0}
\definecolor{COLgrid}{RGB}{220, 220, 220}

\definecolor{COL10}{RGB}{0, 255, 0}
\definecolor{COL50}{RGB}{77, 145, 97}
\definecolor{COL100}{RGB}{255, 166, 0}
\definecolor{COL500}{RGB}{255, 110, 200}
\definecolor{COL1000}{RGB}{66, 233, 255}

\definecolor{COLsdcfr}{RGB}{102, 102, 255}
\definecolor{COLdeepcfr}{RGB}{0, 204, 153}

\usepackage[utf8]{inputenc}
\usepackage[english]{babel}
\usepackage{amsthm}

\theoremstyle{theorem}
\newtheorem{theorem}{Theorem}

\title{Single Deep Counterfactual Regret Minimization}

\author{
  Eric Steinberger\\
  Undergraduate in Computer Science and Technology\\
  University of Cambridge\\
  Cambridge\\
  \texttt{ericsteinberger.est@gmail.com} \\
}

\begin{document}

\maketitle

\begin{abstract}
\textit{Counterfactual Regret Minimization (CFR)} is the most successful algorithm for finding approximate Nash equilibria in \textit{imperfect information games}. However, CFR's reliance on full game-tree traversals limits its scalability and generality. Therefore, the game's state- and action-space is often \textit{abstracted} (i.e. simplified) for CFR, and the resulting strategy is then translated back to the full game. This requires extensive expert-knowledge, is not possible in many games outside of poker, and often converges to highly exploitable policies. A recently proposed method, \textit{Deep CFR}, applies deep learning directly to CFR, allowing the agent to intrinsically abstract and generalize over the state-space from samples, without requiring expert knowledge. In this paper, we introduce \textit{Single Deep CFR (SD-CFR)}, a variant of Deep CFR that has a lower overall approximation error by avoiding the training of an average strategy network. We show that SD-CFR is more attractive from a theoretical perspective and empirically outperforms Deep CFR with respect to exploitability and one-on-one play in poker.
\end{abstract}

\section{Introduction}
In perfect information games, players usually seek to play an optimal deterministic strategy. In contrast, sound policy optimization algorithms for imperfect information games converge towards a \textit{Nash equilibrium}, a distributional strategy characterized by minimizing the losses against a worst-case opponent. The most popular family of algorithms for finding such equilibria is Counterfactual Regret Minimization (CFR) ~\citep{zinkevich2008regret}. Conventional CFR methods iteratively traverse the game-tree to improve the strategy played in each state. For instance, CFR$^+$ ~\citep{CFRp}, a fast variant of CFR, was used to solve two-player Limit Texas Hold'em Poker ~\citep{bowling2015heads, SolvingLH}, a variant of poker frequently played by humans.

However, the scalability of such \textit{tabular} CFR methods is limited since they need to visit a given state to update the policy played in it. In games too large to fully traverse, practitioners hence often employ domain-specific abstraction schemes ~\citep{PotentialAwareAbstr, Habstr} that can be mapped back to the full game after training has finished. Unfortunately, these techniques have been shown to lead to highly exploitable policies in the large benchmark game Heads-Up No-Limit Texas Hold'em Poker (HUNL) ~\citep{lisy2016equilibrium} and typically require extensive expert knowledge to design well. In an attempt to address these two problems, researchers started to augment CFR with neural network function approximation, resulting in DeepStack ~\citep{deepstack}. Concurrently with Libratus ~\citep{Libratus}, DeepStack was one of the first algorithms to defeat professional poker players in HUNL, a game consisting of 10$^{160}$ states and thus being far too large to fully traverse.

While tabular CFR has to visit a state of the game to update its policy in it, a parameterized policy may be able to play an educated strategy in states it has never seen before. Purely parameterized (i.e. non-tabular) policies have led to great breakthroughs in AI for perfect information games ~\citep{Atari, PPO, AlphaGoZero} and were recently also applied to large imperfect information games by Deep CFR ~\citep{brown2018deep} to mimic a variant of tabular CFR from samples.

Deep CFR's strategy relies on a series of two independent neural approximations. In this paper, we introduce \textit{Single Deep CFR (SD-CFR)}, a simplified variant of Deep CFR that obtains its final strategy after just one neural approximation by using what Deep CFR calls \textit{value networks} directly instead of training an additional network to approximate the weighted average strategy. This reduces the overall sampling- and approximation error and makes training more efficient. We show experimentally that SD-CFR improves upon the convergence of Deep CFR in poker games and outperforms Deep CFR in one-one-one matches.

\section{Extensive-form games}
This section introduces extensive-form games and the notation we will use throughout this work. Formally, a finite two-player extensive-form game with imperfect information is a set of \textbf{histories} $\mathcal{H}$, where each history is a path from the root $\phi \in \mathcal{H}$ to any particular state. The subset $\mathcal{Z} \subset \mathcal{H}$ contains all terminal histories. $A(h)$ is the set of actions available to the acting player at history $h$, who is chosen from the set \{1, 2, chance\} by the player function $P(h)$. In any $h \in \mathcal{H}$ where $P(h) = chance$, the action is chosen by the dynamics of the game itself. Let $N = \{1, 2\}$ be the set of both players. When referring to a player $i \in N$, we refer to his opponent by $-i$. All nodes $ z \in \mathcal{Z}$ have an associated \textbf{utility} $u(z)$ for each player. This work focuses on \textbf{zero-sum} games, defined by the property $u_{i}(z) = -u_{-i}(z)$ for all $z \in \mathcal{Z}$.

Imperfect information is represented by partitioning $\mathcal{H}$ into \textbf{information sets}. An information set $I_i$ is a subset of $\mathcal{H}$, where histories $h,h'$ $\in \mathcal{H}$ are in the same information set if and only if player $i$ cannot distinguish between $h$ and $h'$ given his private and all available public information. For each player $i \in N$, an \textbf{information partition} $\mathcal{I}_{i}$ is a set of all such information sets. Let $A(I) = A(h)$ and $P(I) = P(h)$ for all $h \in I$ and each $I \in \mathcal{I}_i$.

Each player $i$ chooses actions according to a \textbf{behavioural strategy} $\sigma_{i}$, with $\sigma_{i}(I, a)$ being the probability of choosing action $a$ when in $I$. We refer to a tuple $(\sigma_1, \sigma_2)$ as a \textbf{strategy profile} $\sigma$. Let $\pi^{\sigma}(h)$ be the probability of reaching history $h$ if both players follow $\sigma$ and let $\pi^{\sigma}_{i}(h)$ be the probability of reaching $h$ if player $i$ acts according to $\sigma_{i}$ and player $-i$ always acts deterministically to get to $h$. It follows that the probability of reaching an information set $I$ if both players follow $\sigma$ is $\pi^{\sigma}(I) = \sum_{h\in I} \pi^{\sigma}(h)$ and is $\pi_i^{\sigma}(I) = \sum_{h\in I} \pi_i^{\sigma}(h)$ if $-i$ plays to get to $I$.

Player $i$'s \textbf{expected utility} from any history $h$ assuming both players follow strategy profile $\sigma$ from $h$ onward is denoted by $u^{\sigma}_{i}(h)$. Thus, their expected utility over the whole game given a strategy profile $\sigma$ can be written as $u^{\sigma}_{i}(\phi) = \sum_{z \in \mathcal{Z}}\pi^{\sigma}(z)u_{i}(z)$.

Finally, a strategy profile $\sigma = (\sigma_{1}, \sigma_{2})$ is a \textbf{Nash equilibrium} if no player $i$ could increase their expected utility by deviating from $\sigma_{i}$ while $-i$ plays according to $\sigma_{-i}$. We measure the \textbf{exploitability} $e(\sigma)$ of a strategy profile by how much its optimal counter strategy profile (also called \textbf{best response}) can beat it by. Let us denote a function that returns the best response to $\sigma_i$ by $BR(\sigma_i)$. Formally,
\begin{equation*}
    e(\sigma) = -\sum_{i \in N} (u_i(\sigma_i, BR(\sigma_i))
\end{equation*}

\section{Counterfactual Regret Minimization (CFR)}
\textit{Counterfactual Regret Minimization (CFR)} ~\citep{zinkevich2008regret} is an iterative algorithm. It can run either \textit{simultaneous} or \textit{alternating} updates. If the former is chosen, CFR produces a new \textit{iteration-strategy} $\sigma_i^t$ for all players $i \in N$ on each iteration $t$. In contrast, alternating updates produce a new strategy for only one player per iteration, with player $t$ $mod$ $2$ updating his on iteration $t$.

To understand how CFR converges to a Nash equilibrium, let us first define the \textit{instantaneous regret} for player $i$ of action $a \in A(I)$ in any $I \in \mathcal{I}_i$ as
\begin{equation} \label{eq:vanCFRimmReg}
    r^t_i(I, a) = \pi_{-i}^{\sigma^t}(I)(v_i^{\sigma^t}(I, a) - v_i^{\sigma^t}(I))
\end{equation}
where \(v_i^{\sigma^t} (I) = \sum_{h \in I} \frac{\pi_{-i}^{\sigma^t}(h) u_i^{\sigma^t}(h)}{\pi_{-i}^{\sigma^t}(I)}\) and \(v_i^{\sigma^t} (I, a) = \sum_{h \in I} \frac{\pi_{-i}^{\sigma^t}(h) u_i^{\sigma^t}(h \xrightarrow{act}a)}{\pi_{-i}^{\sigma^t}(I)}\). Intuitively, $r^t_i(I, a)$ quantifies how much more player $i$ would have won (in expectation), had he always chosen $a$ in $I$ and played to get to $I$ but according to $\sigma^t$ thereafter. The \textit{overall regret} on iteration $T$ is \(R^T_i(I, a) = \sum^T_{t=1} r^t_i(I, a)\). Now, the iteration-strategy for player $i$ can be derived by
\begin{equation} \label{eq:vanCFRiterStrat}
    \sigma^{t+1}_i(I, a) =
    \begin{cases}
        \frac{R^t_i(I, a)_+}{\sum_{\Tilde{a}\in A(I)} R^t_i(I, \Tilde{a})_+}  & \text{if} \sum_{\Tilde{a} \in A(I)} R^t_i(I, \Tilde{a})_+ > 0 \\
        \frac{1}{|A(I)|}                                     & \text{otherwise} \\
    \end{cases}
\end{equation}
where $x_{+} = max(x, 0)$. Note that $\sigma^{0}_i(I, a) = \frac{1}{|A(I)|}$.

The iteration-strategy profile $\sigma^t$ does not converge to an equilibrium as $t \to \infty$ in most variants of CFR \footnote{In CFR-BR ~\citep{CFR_BR} $\sigma^t$ does converge probabilistically as $t\to \inf$ and in CFR$^+$ ~\citep{CFRp} it often does so empirically (but without guarantees); in vanilla CFR and linear CFR ~\citep{LinCFR} $\sigma^t$ typically does not converge.}. The policy that has been shown to converge to an equilibrium profile is the \textit{average strategy} $\bar{\sigma}_i^T$. For all $I \in \mathcal{I}_i$ and each $a \in A(I)$ it is defined as
\begin{equation} \label{eq:vanCFRavgStrat}
    \bar{\sigma}^T_i(I, a) = \frac{\sum_{t=1}^{T} \pi_{i}^{\sigma^t}(I)  \sigma_{i}^{t}(I, a)}{\sum_{t=1}^{T} \pi_{i}^{\sigma^t}(I)}
\end{equation}

\subsection{Variations of CFR}
Aiming to solve ever bigger games, researchers have proposed many improvements upon vanilla CFR over the years ~\citep{SolvingLH, Libratus, deepstack}. These improvements include alternative methods for regret updates ~\citep{CFRp, LinCFR}, automated schemes for abstraction design ~\citep{PotentialAwareAbstr}, and sampling variants of CFR ~\citep{MCCFR}. Many of the most successful algorithms of the recent past also employ \textit{real-time solving} or \textit{re-solving} ~\citep{brown2018depth, deepstack}.

\textit{Discounted CFR (DCFR)} ~\citep{LinCFR} slightly modifies the equations for $R^T_i(I, a)$ and $\bar{\sigma}^T_i$. A special case of DCFR is \textit{linear CFR (LCFR)}, where the contribution of the instantaneous regret of iteration $t$ as well as the contribution of $\sigma^t$ to $\bar{\sigma}^T$ is weighted by $t$. This change alone suffices to let LCFR converge up to two orders of magnitude faster than vanilla CFR does in some large games.

\textit{Monte-Carlo CFR (MC-CFR)} ~\citep{MCCFR} proposes a family of tabular methods that visit only a subset of information sets on each iteration. Different variants of MC-CFR can be constructed by choosing different sampling policies. One such variant is \textit{External Sampling (ES)}, which executes all actions for player $i$, the traverser, in every $I \in \mathcal{I}_i$ but draws only one sample for actions not controlled by $i$ (i.e. those of $-i$ and chance). In games with many player-actions \textit{Average Strategy Sampling} ~\citep{AvrgStratSamplMC}, ~\textit{Robust Sampling} ~\citep{doubleDCFR} are very useful. They, in different ways, sample only a sub-set of actions for $i$. Both LCFR and a similarly fast variant called CFR$^+$ ~\citep{CFRp} are compatible with forms of MC sampling, although CFR$^+$ was regarded as to sensitive to variance until recently ~\citep{VRMCCFR}.

\section{Deep CFR}
CFR methods either need to run on the full game tree or employ domain-specific abstractions. The former is infeasible in large games and the latter not easily possible in all domains. Deep CFR ~\citep{brown2018deep} computes an approximation of linear CFR ~\citep{LinCFR} with alternating player updates. It is sample-based and does not need to store regret tables, making it generally applicable to any two-player zero-sum game.

On each iteration, Deep CFR fits a \textit{value network} $\hat{D_i}$ for one player $i$ to approximate what we call \textit{advantage}, which is defined as \(D_i^T(I, a) = \frac{R^T_{i, linear}(I, a)}{\sum_{t=1}^T (t \pi_{-i}^{\sigma^t}(I))}\), where \(R^T_{i, linear}(I, a) = \sum^T_{t=1} (t r^t_i(I, a))\).

In large games, reach-probabilities naturally are (on average) very small after many tree-branchings. Considering that it is hard for neural networks to learn values across many orders of magnitude ~\citep{NNordersOfMagTargets}, Deep CFR divides $R^T_{i,linear(I, a)}$ by the total linear reach $\sum_{t=1}^T (t \pi_{-i}^{\sigma^t}(I))$ and thereby avoids this problem. This still yields correct results because $\sum_{t=1}^T (t \pi_{-i}^{\sigma^t}(I))$ is identical for all $a \in A(I)$.

We can derive the iteration-strategy for $t+1$ from $D^t$ similarly to CFR in equation \ref{eq:vanCFRiterStrat} by
\begin{equation} \label{eq:DCFRimmStrat}
    \sigma^{t+1}_i(I, a) =
    \begin{cases}
        \frac{D_i^t(I, a)_{+}}{\sum_{\Tilde{a} \in A(I)} D_i^t(I, \Tilde{a})_{+}}  & \text{if} \sum_{\Tilde{a} \in A(I)} D_i^t(I, \Tilde{a})_{+} > 0 \\
        \frac{1}{|A(I)|}           & \text{otherwise} \\
    \end{cases}
\end{equation}

However, Deep CFR modifies this to heuristically choose the action with the highest advantage whenever $\sum_{\Tilde{a} \in A(I)} D_i^t(I, \Tilde{a})_{+} \leq 0$.
Deep CFR obtains the training data for $\hat{D}$ via batched external sampling ~\citep{MCCFR,brown2018deep}. All instantaneous regret values collected over the $N$ traversals are stored in a memory buffer $B^v_i$. After its maximum capacity is reached, $B^v_i$ is updated via reservoir sampling ~\citep{reservoirSampling}. To mimic the behaviour of linear CFR, we need to weight the training losses between the predictions $\hat{D}$ makes and the sampled regret vectors in $B^v_i$ with the iteration-number on which a given datapoint was added to the buffer.

At the end of its training procedure (i.e. after the last iteration), Deep CFR fits another neural network $\hat{S}_i(I, a)$ to approximate the linear average strategy
\begin{equation} \label{eq:DCFRavgStrat}
    \bar{\sigma}^T_i(I, a) = \frac{\sum_{t=1}^{T} (t \pi_{i}^{\sigma^t}(I)  \sigma_{i}^{t}(I, a))} {\sum_{t=1}^{T} (t\pi_{i}^{\sigma^t}(I))}
\end{equation}
Data to train $\hat{S}_i$ is collected in a separate reservoir buffer $B^s_i$ during the same traversals that data for $B^v_i$ is being collected on. Recall that external sampling always samples all actions for the traverser, let us call him $i$, and plays according to $\sigma^t_{-i}$ for the opponent. Thus, when $i$ is the traverser, $-i$ is the one who adds his strategy vector $\sigma_{-i}^t(I)$ to $B^s_{-i}$ in every $I \in \mathcal{I}_{-i}$ visited during this traversal. This elegantly assures that the likelihood of $\sigma_{-i}^{t}(I)$ being added to $B^s_{-i}$ on any given traversal is proportional to $\pi_{-i}^{\sigma^t}(I)$. Like before, we also need to weight the training loss for each datapoint by the iteration-number on which the datapoint was created.

Notice that tabular CFR achieves importance weighting between iterations through multiplying with some form of the reach probability (see equations \ref{eq:vanCFRimmReg} and \ref{eq:vanCFRavgStrat}). In contrast, Deep CFR does so by controlling the expected frequency of datapoints from different iterations occurring in its buffers and by weighting the neural network losses differently for data from each iteration.

\section{Single Deep Counterfactual Regret Minimization (SD-CFR)}
Notice that storing all iteration-strategies would allow one to compute the average strategy on the fly \textit{during play} both in tabular and approximate CFR variants. In tabular methods, the gain of not needing to keep $\bar{\sigma}$ in memory during training would come at the cost of storing $t$ equally large tables (though potentially on disk) during training and during play. However, this is very different with Deep CFR. Not aggregating into $\hat{S}$ removes the sampling- and approximation error that $B^s$ and $\hat{S}$ introduce, respectively. Moreover, the computational work needed to train $\hat{S}$ is no longer required. Like in the tabular case, we do need to keep all iteration strategies, but this is much cheaper with Deep CFR as strategies are compressed within small neural networks.

We will now look at two methods for querying $\bar{\sigma}$ from a buffer of past value networks $B^M$.

\subsection{Acting on freely playable trajectories}
Often (e.g. in one-one-one evaluations and during rollouts), a trajectory is played from the root of the game-tree and the agent is only required to return action-samples of the average strategy on each step forward. In this case, SD-CFR chooses a value network $\hat{D}^t_i \in B_i^M$ at the start of the game, where each $\hat{D}^t_i$ is assigned sampling weight $t$. The policy $\sigma_i$, which this network gives by equation \ref{eq:DCFRimmStrat}, is now going to be used for the whole game trajectory. We call this method \textit{trajectory-sampling}.

By applying the sampling weights when selecting a $\hat{D}_i \in B_i^M$, we satisfy the linear averaging constraint of equation \ref{eq:DCFRavgStrat}, and by using the same $\sigma_i$ for the whole trajectory starting at the root, we ensure that the iteration-strategies are also weighted proportionally to each of their reach-probabilities in any given state along that trajectory. The latter happens naturally, since $\hat{D^t_i}$ of any $t$ produces $\sigma_i^t$, which reaches each information set $I$ with a likelihood directly proportional to $\pi_i^{\sigma^t}(I)$ when playing from the root.

The query cost of this method is constant with the number of iterations (and equal to the cost of querying Deep CFR).

\subsection{Querying a complete action distribution in any information set}
Let us now consider querying the complete action probability distribution $\bar{\sigma}^T_i(I)$ in some information set $I \in \mathcal{I}_i$. Given $B^M_i$, we can compute $\bar{\sigma}^T_i(I)$ exactly through equation \ref{eq:DCFRavgStrat}, where we compute
\begin{equation}
    \pi_i^{\sigma^t}(I) = \prod_{I' \in I, P(I')=i, a': I'\to I} \sigma^{t}_i(I', a')
\end{equation}
Here, $I' \in I$ means that $I'$ is on the trajectory leading to $I$ and $a': I'\to I$ is the action selected in $I'$ leading to $I$.

This computation can be done with at most\footnote{This number can further be reduced by omitting queries for any $\sigma^t$ as soon as it assigns probability 0 to the action played on the trajectory.} as many feedforward passes through each network in $B^M_i$ as player $i$ had decisions along the trajectory to $I$, typically taking a few seconds in poker when done on a CPU.

\subsection{Querying a complete action distribution on a trajectory}
If a trajectory is played forward from the root, as is the case in e.g. exploitability evaluation, we can cache the step-wise reach-probabilities on each step $I^k$ along the trajectory and compute $\pi_i^{\sigma^t}(I^{k+1}) = \sigma_i^t(I^{k+1}, a') \pi_i^{\sigma^t}(I^k)$, where $a'$ is the action that leads from $I^k$ to $I^{k+1}$. This reduces the number of queries per step to at most $|B^M_i|$.

\subsection{Theoretical and practical properties}
SD-CFR always mimics $\bar{\sigma}^T_i$ correctly from the iteration-strategies it is given. Thus, if these iteration-strategies were perfect approximations of the real iteration-strategies, SD-CFR is equivalent to linear CFR (see Theorem \ref{theorem:SDisSound}), which is not necessarily true for Deep CFR (see Theorem \ref{theorem:FiniteBuffer}).

As we later show in an experiment, SD-CFR's performance degrades if reservoir sampling is performed on $B^M$ after the number of iterations trained exceeds the buffer's capacity. Thankfully, the neural network proposed to be used for Deep CFR in large poker games has under 100,000 parameters ~\citep{brown2018deep} and thus requires under 400KB of disk space. Deep CFR is usually trained for just a few hundred iterations ~\citep{brown2018deep}, but storing even 25,000 such networks on disk would need only 10GB of disk space. At no point during any computation do we need all networks in memory. Thus, keeping all value networks will not represent a problem in practise.

Observing that Deep CFR and SD-CFR depend upon the accuracy of the value networks in exactly the same way, we can conclude that SD-CFR is a better or equally good approximation of linear CFR as long as all value networks are stored. Though this shows that SD-CFR is largely superior in theory, it is not implicit that SD-CFR will always produce stronger strategies empirically. We will investigate this next.

\begin{theorem} \label{theorem:FiniteBuffer}
    If the capacity of strategy buffer $B_i^s$ is finite or if only a finite number K of traversals is executed per iteration, $B_i^s$ is not guaranteed to reflect the true average strategy $\bar{\sigma}^T_i(I)$ for every $I \in \mathcal{I}_i$ even if all value networks are perfect approximators of the true advantage after any number of training iterations $T>2$. Hence, even a perfect function approximator for $\hat{S}$ is not guaranteed to model $\bar{\sigma}^T_i$ without error.
\end{theorem}

\begin{theorem} \label{theorem:SDisSound}
    Assume that for all $i \in N$, all $I \in \mathcal{I}_i$, all $a \in A(I)$, and all $t$ up to the number of iterations trained $T$, $\hat{D}_i^t(I, a) = D_i^t(I, a)$ (i.e. that all value networks perfectly model the true advantages). Now, SD-CFR represents $\bar{\sigma}^T_i$ without error. This holds for both trajectory-sampling SD-CFR and for when SD-CFR computes $\bar{\sigma}^T_i(I)$ explicitly. Furthermore, an opponent has no way of distinguishing which of the two proposed methods of sampling from $\bar{\sigma}$ is used solely from gameplay.
\end{theorem}

Proofs for both Theorem \ref{theorem:FiniteBuffer} and \ref{theorem:SDisSound} can be found in the appendix.

\section{Experiments}
We empirically evaluate SD-CFR by comparing to Deep CFR and by analyzing the effect of sampling on $B^M$. Recall that Deep CFR and SD-CFR are equivalent in how they train their value networks. This allows both algorithms to \textit{share the same value networks} in our experiments, which makes comparisons far less susceptible to variance over algorithm runs and conveniently guarantees that both algorithms tend to the same Nash equilibrium.

Where not otherwise noted, we use hyperparamters as ~\cite{brown2018deep} do. Our environment observations include additional features such as the size of the pot and represent cards as concatenated one-hot vectors without any higher level features, but are otherwise as ~\cite{brown2018deep}.

\subsection{Leduc}
\begin{figure}[H]
\begin{center}
\captionsetup[subfigure]{width=0.38\linewidth}
\subfloat[Comparing SD-CFR and Deep CFR]
{   \label{fig:explLeduc}
        \begin{tikzpicture}[scale=0.75]
        \pgfplotsset{
            legend style={font=\footnotesize}
            }
            \begin{loglogaxis}[
                xlabel={Algorithm Iterations},
                ylabel={Exploitability in mA/g},
                xmin=10, xmax=5010,
                ymin=30, ymax=1070,
                legend pos=north east,
                legend cell align={left},
                ymajorgrids=true,
                grid style={thin, solid, COLgrid},
            ]
            
            \addplot[
                name path=DeepCFR,
                color=COLdeepcfr, 
                style={ultra thick},
                solid,
                mark options={solid},
                ] coordinates 
           {(1.00000,2329.87725)(30.00000,401.36666)(60.00000,230.35985)(90.00000,199.06578)(120.00000,169.65290)(150.00000,148.17958)(180.00000,151.25171)(210.00000,155.24555)(240.00000,145.33548)(270.00000,134.95342)(300.00000,142.95615)(330.00000,125.20101)(360.00000,133.44337)(390.00000,142.57031)(420.00000,135.29160)(450.00000,122.75385)(480.00000,117.50766)(510.00000,115.19369)(540.00000,125.45157)(570.00000,116.00748)(600.00000,113.73451)(630.00000,122.68163)(660.00000,122.82616)(690.00000,124.90866)(720.00000,124.20836)(750.00000,118.29062)(780.00000,112.17643)(810.00000,115.29979)(840.00000,110.08565)(870.00000,119.96120)(900.00000,113.24303)(930.00000,108.05143)(960.00000,112.45266)(990.00000,105.71928)(1020.00000,115.68286)(1050.00000,109.58418)(1080.00000,109.41454)(1110.00000,102.90915)(1140.00000,116.99124)(1170.00000,105.23453)(1200.00000,110.87190)(1230.00000,114.81875)(1260.00000,107.26768)(1290.00000,106.32132)(1320.00000,106.01863)(1350.00000,112.97851)(1380.00000,116.87868)(1410.00000,105.76573)(1440.00000,105.17453)(1470.00000,96.63560)(1500.00000,95.08351)(1530.00000,100.20406)(1560.00000,109.16933)(1590.00000,108.73039)(1620.00000,95.66571)(1650.00000,100.25539)(1680.00000,86.65326)(1710.00000,94.44984)(1740.00000,99.87646)(1770.00000,116.81765)(1800.00000,96.71766)(1830.00000,101.40729)(1860.00000,99.54209)(1890.00000,98.20840)(1920.00000,95.18048)(1950.00000,90.05814)(1980.00000,89.16577)(2010.00000,93.83230)(2040.00000,87.92884)(2070.00000,92.69410)(2100.00000,90.89312)(2130.00000,100.33958)(2160.00000,89.29004)(2190.00000,87.98694)(2220.00000,95.43586)(2250.00000,90.09825)(2280.00000,86.84998)(2310.00000,104.46784)(2340.00000,82.72936)(2370.00000,93.33342)(2400.00000,84.64486)(2430.00000,86.24474)(2460.00000,97.87968)(2490.00000,91.10215)(2520.00000,110.47056)(2550.00000,94.20192)(2580.00000,85.16060)(2610.00000,87.76592)(2640.00000,85.53110)(2670.00000,93.03817)(2700.00000,98.49396)(2730.00000,103.78026)(2760.00000,92.24077)(2790.00000,88.20704)(2820.00000,87.80089)(2850.00000,95.13255)(2880.00000,92.28657)(2910.00000,97.68215)(2940.00000,98.59008)(2970.00000,93.99571)(3000.00000,85.78455)(3030.00000,91.17435)(3060.00000,94.64639)(3090.00000,91.89557)(3120.00000,83.02606)(3150.00000,84.85731)(3180.00000,90.53113)(3210.00000,88.58172)(3240.00000,83.90681)(3270.00000,79.63277)(3300.00000,83.48500)(3330.00000,84.03857)(3360.00000,85.41210)(3390.00000,84.47598)(3420.00000,86.09435)(3450.00000,89.32511)(3480.00000,80.52412)(3510.00000,82.13733)(3540.00000,81.97733)(3570.00000,79.59257)(3600.00000,82.85706)(3630.00000,74.21497)(3660.00000,83.06617)(3690.00000,84.23992)(3720.00000,84.53271)(3750.00000,89.86126)(3780.00000,80.65975)(3810.00000,87.90810)(3840.00000,78.85295)(3870.00000,78.92248)(3900.00000,84.67244)(3930.00000,81.23492)(3960.00000,74.48831)(3990.00000,82.42862)(4020.00000,77.40400)(4050.00000,80.23218)(4080.00000,84.25324)(4110.00000,86.90082)(4140.00000,83.15259)(4170.00000,88.00695)(4200.00000,83.05820)(4230.00000,88.52725)(4260.00000,85.30708)(4290.00000,84.47292)(4320.00000,84.75630)(4350.00000,84.89510)(4380.00000,80.25246)(4410.00000,83.56138)(4440.00000,83.15900)(4470.00000,83.79317)(4500.00000,90.12319)(4530.00000,77.45139)(4560.00000,81.97434)(4590.00000,86.35762)(4620.00000,89.90510)(4650.00000,79.66766)(4680.00000,77.78592)(4710.00000,79.11505)(4740.00000,83.19963)(4770.00000,88.02776)(4800.00000,84.64974)(4830.00000,82.48286)(4860.00000,81.66603)(4890.00000,83.03196)(4920.00000,86.13944)(4950.00000,78.52865)(4980.00000,79.86621)};

            \addlegendentry{Deep CFR}
            
            \addplot[
                name path=SDCFR,
                color=COLsdcfr,
                style={ultra thick},
                solid,
                mark options={solid},
                ] coordinates
            {(1.00000,2373.61133)(30.00000,381.45848)(60.00000,226.78933)(90.00000,186.00441)(120.00000,154.39717)(150.00000,141.66960)(180.00000,141.68800)(210.00000,138.73936)(240.00000,126.47322)(270.00000,116.73802)(300.00000,111.71467)(330.00000,110.83163)(360.00000,116.82523)(390.00000,117.34464)(420.00000,110.26822)(450.00000,111.19175)(480.00000,102.92437)(510.00000,94.61301)(540.00000,94.57735)(570.00000,93.64876)(600.00000,92.58759)(630.00000,97.06905)(660.00000,102.77252)(690.00000,104.18112)(720.00000,98.43136)(750.00000,99.75137)(780.00000,95.91319)(810.00000,91.96688)(840.00000,89.13601)(870.00000,91.49228)(900.00000,89.94055)(930.00000,86.22172)(960.00000,92.25098)(990.00000,88.66829)(1020.00000,88.72790)(1050.00000,89.12872)(1080.00000,88.28701)(1110.00000,86.69406)(1140.00000,87.72501)(1170.00000,86.91777)(1200.00000,87.29684)(1230.00000,86.57532)(1260.00000,88.57685)(1290.00000,88.47544)(1320.00000,86.70290)(1350.00000,87.87214)(1380.00000,83.61359)(1410.00000,84.16934)(1440.00000,82.24647)(1470.00000,78.68258)(1500.00000,74.60876)(1530.00000,74.69924)(1560.00000,76.55276)(1590.00000,79.35901)(1620.00000,78.04755)(1650.00000,73.96620)(1680.00000,74.43080)(1710.00000,76.58732)(1740.00000,80.31264)(1770.00000,79.68849)(1800.00000,79.13798)(1830.00000,77.16058)(1860.00000,74.77252)(1890.00000,72.96379)(1920.00000,71.72141)(1950.00000,69.74870)(1980.00000,69.87396)(2010.00000,69.30000)(2040.00000,68.46460)(2070.00000,67.87156)(2100.00000,68.05638)(2130.00000,66.97242)(2160.00000,67.70881)(2190.00000,70.32812)(2220.00000,71.32339)(2250.00000,69.84151)(2280.00000,67.86035)(2310.00000,66.97885)(2340.00000,67.61827)(2370.00000,68.50820)(2400.00000,68.02876)(2430.00000,67.91904)(2460.00000,67.75500)(2490.00000,66.82547)(2520.00000,65.76090)(2550.00000,64.77975)(2580.00000,65.79037)(2610.00000,66.23838)(2640.00000,65.03359)(2670.00000,67.64484)(2700.00000,66.96325)(2730.00000,66.37604)(2760.00000,66.63423)(2790.00000,66.62085)(2820.00000,67.59369)(2850.00000,68.70652)(2880.00000,68.26175)(2910.00000,67.74465)(2940.00000,68.05917)(2970.00000,67.21598)(3000.00000,67.27689)(3030.00000,66.70341)(3060.00000,65.48807)(3090.00000,65.87108)(3120.00000,65.48552)(3150.00000,63.23693)(3180.00000,65.03636)(3210.00000,64.85020)(3240.00000,63.72582)(3270.00000,62.89599)(3300.00000,62.11175)(3330.00000,61.56929)(3360.00000,61.08648)(3390.00000,61.25096)(3420.00000,61.65993)(3450.00000,62.00178)(3480.00000,61.78718)(3510.00000,61.38363)(3540.00000,60.40469)(3570.00000,59.38618)(3600.00000,58.32509)(3630.00000,58.71773)(3660.00000,59.46145)(3690.00000,59.93789)(3720.00000,60.57844)(3750.00000,60.66835)(3780.00000,59.72911)(3810.00000,59.94697)(3840.00000,60.55234)(3870.00000,60.37849)(3900.00000,60.20205)(3930.00000,59.57906)(3960.00000,58.71832)(3990.00000,58.68466)(4020.00000,58.67017)(4050.00000,59.18573)(4080.00000,60.37195)(4110.00000,60.83474)(4140.00000,61.12495)(4170.00000,61.44301)(4200.00000,61.19049)(4230.00000,60.16495)(4260.00000,59.78807)(4290.00000,59.99441)(4320.00000,59.26126)(4350.00000,58.99326)(4380.00000,59.14012)(4410.00000,58.23881)(4440.00000,57.67924)(4470.00000,57.02022)(4500.00000,57.63453)(4530.00000,57.81806)(4560.00000,58.09892)(4590.00000,58.21695)(4620.00000,57.47830)(4650.00000,57.88988)(4680.00000,57.76441)(4710.00000,57.96801)(4740.00000,58.38671)(4770.00000,58.90518)(4800.00000,59.51113)(4830.00000,59.86625)(4860.00000,59.79977)(4890.00000,59.53355)(4920.00000,58.91913)(4950.00000,58.91983)(4980.00000,59.02260)};
            
            \addlegendentry{SD-CFR (ours)}
        
            \addplot[mark=none, COLZeroLine, samples=2222, forget plot] coordinates {
            (0,  0)
            (10000,  0)
            };
            
            \end{loglogaxis}
\end{tikzpicture}
}
\subfloat[Sampling on $B^M$ with finite capacity]
{   \label{fig:explResSampl}
       \begin{tikzpicture}[scale=0.75]
        \pgfplotsset{
            legend style={font=\tiny, fill opacity=0.92, draw opacity=1, text opacity =1},
            legend image code/.code={
                \draw[mark repeat=2,mark phase=2]
                plot coordinates {
                (0cm,0cm)
                (0.15cm,0cm)       
                (0.3cm,0cm)         
                };%
            }
        }   
            \begin{loglogaxis}[
                legend cell align={left},
                xlabel={Algorithm Iterations},
                xmin=10, xmax=5010,
                ymin=30, ymax=1070,
                legend pos=south west,
                ymajorgrids=true,
                grid style={thin, solid, COLgrid},
            ]
            
            \addlegendimage{empty legend}
            \addlegendentry{\hspace{-.23mm}Capacity of $B^M$}
            \addplot[
                name path=10,
                color=COL10, 
                style={ultra thick},
                solid,
                mark options={solid},
                ] coordinates 
            {(1.00000,2373.61133)(30.00000,447.20458)(60.00000,343.77701)(90.00000,322.85775)(120.00000,313.10910)(150.00000,316.86789)(180.00000,385.46416)(210.00000,365.20710)(240.00000,411.59892)(270.00000,402.71922)(300.00000,400.96690)(330.00000,412.35120)(360.00000,343.31402)(390.00000,347.11889)(420.00000,365.16192)(450.00000,400.96691)(480.00000,465.58167)(510.00000,468.37308)(540.00000,446.00084)(570.00000,446.00084)(600.00000,425.23997)(630.00000,484.30959)(660.00000,485.20079)(690.00000,500.33468)(720.00000,523.32799)(750.00000,542.00127)(780.00000,525.71207)(810.00000,566.51777)(840.00000,566.51777)(870.00000,571.24734)(900.00000,543.32713)(930.00000,543.32713)(960.00000,541.39549)(990.00000,499.68567)(1020.00000,517.04906)(1050.00000,561.41295)(1080.00000,602.23941)(1110.00000,635.71872)(1140.00000,635.71872)(1170.00000,642.18773)(1200.00000,642.18773)(1230.00000,601.78629)(1260.00000,541.53413)(1290.00000,541.53413)(1320.00000,500.84444)(1350.00000,500.84479)(1380.00000,500.84479)(1410.00000,500.84479)(1440.00000,465.57218)(1470.00000,465.57218)(1500.00000,429.03360)(1530.00000,380.16960)(1560.00000,399.38926)(1590.00000,425.09298)(1620.00000,439.17173)(1650.00000,439.17173)(1680.00000,439.17173)(1710.00000,440.48794)(1740.00000,452.38878)(1770.00000,416.05157)(1800.00000,416.05157)(1830.00000,408.63137)(1860.00000,408.63137)(1890.00000,408.63137)(1920.00000,410.59792)(1950.00000,410.59792)(1980.00000,415.79544)(2010.00000,434.15768)(2040.00000,432.03598)(2070.00000,476.87382)(2100.00000,476.87382)(2130.00000,488.24768)(2160.00000,496.88549)(2190.00000,496.88549)(2220.00000,496.88549)(2250.00000,525.91176)(2280.00000,525.91176)(2310.00000,525.91176)(2340.00000,549.57702)(2370.00000,546.79343)(2400.00000,546.79343)(2430.00000,582.26381)(2460.00000,582.26381)(2490.00000,582.26381)(2520.00000,582.26381)(2550.00000,583.83315)(2580.00000,574.76056)(2610.00000,574.76056)(2640.00000,574.76056)(2670.00000,574.76056)(2700.00000,535.98703)(2730.00000,550.45869)(2760.00000,550.45869)(2790.00000,550.45869)(2820.00000,550.45869)(2850.00000,545.62956)(2880.00000,545.62956)(2910.00000,545.62956)(2940.00000,545.62956)(2970.00000,545.62956)(3000.00000,532.64666)(3030.00000,532.64666)(3060.00000,552.43103)(3090.00000,552.43103)(3120.00000,552.43103)(3150.00000,552.43103)(3180.00000,552.43103)(3210.00000,562.21906)(3240.00000,562.21906)(3270.00000,522.23876)(3300.00000,522.23876)(3330.00000,522.23876)(3360.00000,522.23876)(3390.00000,546.29948)(3420.00000,546.29948)(3450.00000,546.29948)(3480.00000,546.29948)(3510.00000,546.29948)(3540.00000,546.29948)(3570.00000,546.29948)(3600.00000,602.60641)(3630.00000,602.60641)(3660.00000,602.60641)(3690.00000,635.80954)(3720.00000,635.80954)(3750.00000,606.54599)(3780.00000,606.54599)(3810.00000,600.72274)(3840.00000,600.72274)(3870.00000,600.72274)(3900.00000,600.72274)(3930.00000,597.68477)(3960.00000,597.68477)(3990.00000,597.68477)(4020.00000,597.68477)(4050.00000,605.22982)(4080.00000,593.29340)(4110.00000,593.29340)(4140.00000,567.07928)(4170.00000,567.35313)(4200.00000,567.35313)(4230.00000,567.35313)(4260.00000,570.04335)(4290.00000,570.04335)(4320.00000,570.04335)(4350.00000,570.04335)(4380.00000,570.04335)(4410.00000,570.04335)(4440.00000,590.57244)(4470.00000,591.69825)(4500.00000,591.69825)(4530.00000,574.72873)(4560.00000,574.72873)(4590.00000,601.28231)(4620.00000,576.15706)(4650.00000,576.15706)(4680.00000,576.15706)(4710.00000,569.20211)(4740.00000,569.20211)(4770.00000,569.20211)(4800.00000,545.92190)(4830.00000,545.92190)(4860.00000,545.92190)(4890.00000,545.92190)(4920.00000,545.92190)(4950.00000,541.80151)(4980.00000,541.80151)(5010.00000,508.87978)};
            \addlegendentry{10 (4MiB)}
            
            \addplot[
                name path=50,
                color=COL50, 
                style={ultra thick},
                solid,
                mark options={solid},
                ] coordinates 
           {(1.00000,2373.61133)(30.00000,333.72956)(60.00000,178.25092)(90.00000,169.38950)(120.00000,164.47572)(150.00000,153.67655)(180.00000,160.31435)(210.00000,155.94273)(240.00000,182.06424)(270.00000,168.46850)(300.00000,137.15530)(330.00000,147.54546)(360.00000,155.98034)(390.00000,176.45411)(420.00000,189.48666)(450.00000,192.83567)(480.00000,182.47792)(510.00000,182.86236)(540.00000,189.58981)(570.00000,178.97640)(600.00000,197.50864)(630.00000,206.34223)(660.00000,187.30998)(690.00000,187.43095)(720.00000,206.77673)(750.00000,228.53443)(780.00000,226.25611)(810.00000,207.67853)(840.00000,202.65558)(870.00000,209.55930)(900.00000,222.48327)(930.00000,206.60902)(960.00000,196.32488)(990.00000,187.94381)(1020.00000,171.42238)(1050.00000,180.10669)(1080.00000,188.78461)(1110.00000,199.91043)(1140.00000,204.25601)(1170.00000,186.31585)(1200.00000,186.80554)(1230.00000,193.97597)(1260.00000,195.69372)(1290.00000,209.28572)(1320.00000,209.96976)(1350.00000,213.23955)(1380.00000,216.71932)(1410.00000,205.31170)(1440.00000,191.39763)(1470.00000,189.83866)(1500.00000,182.88968)(1530.00000,181.78175)(1560.00000,191.74427)(1590.00000,193.47667)(1620.00000,212.66290)(1650.00000,195.08947)(1680.00000,178.22145)(1710.00000,180.71733)(1740.00000,189.75471)(1770.00000,201.26309)(1800.00000,207.41913)(1830.00000,210.95054)(1860.00000,199.93399)(1890.00000,188.80189)(1920.00000,198.61227)(1950.00000,188.38324)(1980.00000,194.64670)(2010.00000,185.06667)(2040.00000,183.22131)(2070.00000,172.39787)(2100.00000,175.74786)(2130.00000,172.30442)(2160.00000,172.30442)(2190.00000,160.32080)(2220.00000,163.58934)(2250.00000,160.78601)(2280.00000,171.90967)(2310.00000,165.94476)(2340.00000,165.43472)(2370.00000,170.22675)(2400.00000,169.03411)(2430.00000,174.28771)(2460.00000,182.45348)(2490.00000,177.92466)(2520.00000,171.33700)(2550.00000,162.93596)(2580.00000,165.86910)(2610.00000,183.64734)(2640.00000,185.39264)(2670.00000,187.35483)(2700.00000,180.68717)(2730.00000,189.27075)(2760.00000,189.45458)(2790.00000,189.45458)(2820.00000,187.21801)(2850.00000,185.26968)(2880.00000,179.54533)(2910.00000,174.44219)(2940.00000,174.61923)(2970.00000,185.10802)(3000.00000,193.93190)(3030.00000,198.22030)(3060.00000,200.24522)(3090.00000,200.33422)(3120.00000,199.59069)(3150.00000,198.24002)(3180.00000,194.61358)(3210.00000,198.09781)(3240.00000,198.61121)(3270.00000,201.57140)(3300.00000,206.69338)(3330.00000,210.01757)(3360.00000,224.57012)(3390.00000,216.55320)(3420.00000,222.95506)(3450.00000,231.57195)(3480.00000,232.55338)(3510.00000,224.11361)(3540.00000,219.41984)(3570.00000,218.28352)(3600.00000,218.98042)(3630.00000,221.11172)(3660.00000,223.27392)(3690.00000,225.21840)(3720.00000,220.36574)(3750.00000,216.11829)(3780.00000,216.11829)(3810.00000,214.63624)(3840.00000,212.03191)(3870.00000,212.03191)(3900.00000,211.91126)(3930.00000,210.67382)(3960.00000,210.67382)(3990.00000,213.44398)(4020.00000,211.99531)(4050.00000,214.46582)(4080.00000,208.59191)(4110.00000,199.12929)(4140.00000,198.75919)(4170.00000,198.53774)(4200.00000,209.18071)(4230.00000,215.17633)(4260.00000,228.76064)(4290.00000,242.26936)(4320.00000,249.33356)(4350.00000,248.69407)(4380.00000,247.08137)(4410.00000,241.52793)(4440.00000,241.25293)(4470.00000,236.17333)(4500.00000,237.57139)(4530.00000,232.74068)(4560.00000,227.89263)(4590.00000,220.62134)(4620.00000,219.08803)(4650.00000,224.96564)(4680.00000,229.26615)(4710.00000,217.74214)(4740.00000,211.40255)(4770.00000,208.12248)(4800.00000,211.85297)(4830.00000,216.90239)(4860.00000,212.52900)(4890.00000,208.51379)(4920.00000,207.07372)(4950.00000,199.27710)(4980.00000,199.47691)(5010.00000,199.47691)};

            \addlegendentry{50 (20MiB)}
            
            \addplot[
                name path=100,
                color=COL100,
                style={ultra thick},
                solid,
                mark options={solid},
                ] coordinates
           {(1.00000,2373.61133)(30.00000,419.29386)(60.00000,228.28826)(90.00000,170.79908)(120.00000,158.39629)(150.00000,136.90407)(180.00000,134.25759)(210.00000,123.65135)(240.00000,122.29921)(270.00000,135.84303)(300.00000,130.46069)(330.00000,114.73308)(360.00000,116.26617)(390.00000,122.79994)(420.00000,142.12574)(450.00000,139.49300)(480.00000,118.25384)(510.00000,123.01711)(540.00000,121.91334)(570.00000,102.41162)(600.00000,121.63794)(630.00000,132.20693)(660.00000,138.95328)(690.00000,144.48798)(720.00000,126.17296)(750.00000,126.48351)(780.00000,140.85858)(810.00000,149.89101)(840.00000,125.56447)(870.00000,125.26908)(900.00000,142.90096)(930.00000,144.97349)(960.00000,120.62751)(990.00000,117.14212)(1020.00000,130.60673)(1050.00000,134.57898)(1080.00000,120.01627)(1110.00000,119.00904)(1140.00000,118.76449)(1170.00000,122.18434)(1200.00000,133.12841)(1230.00000,156.91891)(1260.00000,155.89996)(1290.00000,150.67053)(1320.00000,138.51040)(1350.00000,142.31029)(1380.00000,157.53141)(1410.00000,171.02701)(1440.00000,172.04340)(1470.00000,152.72013)(1500.00000,142.07721)(1530.00000,146.09589)(1560.00000,145.38211)(1590.00000,156.25510)(1620.00000,161.58975)(1650.00000,156.02320)(1680.00000,152.03851)(1710.00000,139.26744)(1740.00000,137.51254)(1770.00000,129.41394)(1800.00000,128.77262)(1830.00000,131.07860)(1860.00000,130.51816)(1890.00000,130.65488)(1920.00000,137.20091)(1950.00000,130.87281)(1980.00000,126.18229)(2010.00000,118.77757)(2040.00000,119.00125)(2070.00000,118.15208)(2100.00000,111.94066)(2130.00000,113.99456)(2160.00000,114.37128)(2190.00000,109.54725)(2220.00000,111.49297)(2250.00000,113.67257)(2280.00000,118.17205)(2310.00000,118.56922)(2340.00000,121.88152)(2370.00000,118.63590)(2400.00000,125.00502)(2430.00000,129.77822)(2460.00000,128.93976)(2490.00000,124.47329)(2520.00000,118.42543)(2550.00000,118.00967)(2580.00000,121.72466)(2610.00000,121.40863)(2640.00000,121.12953)(2670.00000,125.39896)(2700.00000,128.44673)(2730.00000,127.47440)(2760.00000,136.46911)(2790.00000,139.91220)(2820.00000,142.16531)(2850.00000,140.90295)(2880.00000,140.02781)(2910.00000,133.30761)(2940.00000,138.50603)(2970.00000,139.39115)(3000.00000,132.83317)(3030.00000,126.24513)(3060.00000,123.22647)(3090.00000,118.55245)(3120.00000,114.28358)(3150.00000,115.03632)(3180.00000,115.32655)(3210.00000,112.47599)(3240.00000,110.26234)(3270.00000,113.82459)(3300.00000,118.02466)(3330.00000,109.81081)(3360.00000,105.06946)(3390.00000,110.67056)(3420.00000,107.96198)(3450.00000,104.61814)(3480.00000,102.10876)(3510.00000,105.33465)(3540.00000,113.06001)(3570.00000,115.79035)(3600.00000,115.74231)(3630.00000,120.63587)(3660.00000,121.16477)(3690.00000,120.71732)(3720.00000,119.98673)(3750.00000,113.55939)(3780.00000,119.27581)(3810.00000,113.52957)(3840.00000,116.44949)(3870.00000,115.40251)(3900.00000,115.63514)(3930.00000,113.92431)(3960.00000,109.92975)(3990.00000,107.84482)(4020.00000,108.54949)(4050.00000,104.89096)(4080.00000,104.91407)(4110.00000,102.40560)(4140.00000,100.94172)(4170.00000,100.76240)(4200.00000,101.02846)(4230.00000,101.74896)(4260.00000,101.74644)(4290.00000,109.52374)(4320.00000,111.49474)(4350.00000,111.91815)(4380.00000,113.37092)(4410.00000,113.69501)(4440.00000,111.35338)(4470.00000,118.23748)(4500.00000,119.42611)(4530.00000,118.51654)(4560.00000,120.63529)(4590.00000,122.60061)(4620.00000,119.23350)(4650.00000,119.46346)(4680.00000,115.85727)(4710.00000,117.76838)(4740.00000,114.27459)(4770.00000,117.98275)(4800.00000,114.10599)(4830.00000,117.72951)(4860.00000,119.13401)(4890.00000,121.60576)(4920.00000,117.80319)(4950.00000,115.47095)(4980.00000,116.69895)(5010.00000,117.90517)};
            \addlegendentry{100 (40MiB)}

            \addplot[
                name path=500,
                color=COL500,
                style={ultra thick},
                solid,
                mark options={solid},
                ] coordinates
           {(1.00000,2373.61133)(30.00000,343.07799)(60.00000,196.42673)(90.00000,148.61498)(120.00000,141.36739)(150.00000,143.76777)(180.00000,134.28795)(210.00000,126.44751)(240.00000,124.53562)(270.00000,110.75267)(300.00000,107.77792)(330.00000,102.41958)(360.00000,105.83851)(390.00000,119.43117)(420.00000,116.28886)(450.00000,109.61626)(480.00000,101.59561)(510.00000,99.76410)(540.00000,97.31061)(570.00000,90.46079)(600.00000,98.40257)(630.00000,107.84066)(660.00000,106.65913)(690.00000,114.24556)(720.00000,111.43610)(750.00000,102.97826)(780.00000,100.24553)(810.00000,98.61398)(840.00000,101.81752)(870.00000,101.66777)(900.00000,105.93283)(930.00000,105.98944)(960.00000,106.95590)(990.00000,104.02520)(1020.00000,97.58566)(1050.00000,96.80105)(1080.00000,94.75306)(1110.00000,94.65715)(1140.00000,96.72512)(1170.00000,96.29680)(1200.00000,93.89847)(1230.00000,97.96423)(1260.00000,97.70059)(1290.00000,95.98922)(1320.00000,95.29910)(1350.00000,95.59414)(1380.00000,92.56139)(1410.00000,86.36717)(1440.00000,84.63671)(1470.00000,85.83147)(1500.00000,84.53851)(1530.00000,82.22265)(1560.00000,80.85976)(1590.00000,79.58948)(1620.00000,79.52761)(1650.00000,79.56620)(1680.00000,79.78232)(1710.00000,77.79802)(1740.00000,79.30102)(1770.00000,80.33617)(1800.00000,79.75114)(1830.00000,84.27661)(1860.00000,89.11504)(1890.00000,89.34110)(1920.00000,91.06174)(1950.00000,93.48886)(1980.00000,98.66238)(2010.00000,97.92859)(2040.00000,107.38266)(2070.00000,112.60977)(2100.00000,114.91484)(2130.00000,119.53412)(2160.00000,116.88620)(2190.00000,117.98111)(2220.00000,111.83337)(2250.00000,109.45887)(2280.00000,105.99336)(2310.00000,103.71204)(2340.00000,99.21010)(2370.00000,94.90881)(2400.00000,93.87776)(2430.00000,93.17838)(2460.00000,93.45990)(2490.00000,91.03163)(2520.00000,91.81275)(2550.00000,95.24333)(2580.00000,98.22187)(2610.00000,101.24536)(2640.00000,102.00134)(2670.00000,99.39474)(2700.00000,98.89257)(2730.00000,95.94718)(2760.00000,94.32582)(2790.00000,94.04984)(2820.00000,91.01028)(2850.00000,88.00995)(2880.00000,88.41104)(2910.00000,88.17252)(2940.00000,87.21093)(2970.00000,85.93908)(3000.00000,86.28233)(3030.00000,84.83182)(3060.00000,82.02048)(3090.00000,83.96036)(3120.00000,86.14204)(3150.00000,86.98806)(3180.00000,87.64452)(3210.00000,89.74026)(3240.00000,90.58189)(3270.00000,90.20980)(3300.00000,88.77621)(3330.00000,89.80275)(3360.00000,85.89995)(3390.00000,83.82025)(3420.00000,81.68667)(3450.00000,82.88527)(3480.00000,82.16228)(3510.00000,81.43069)(3540.00000,82.96240)(3570.00000,86.38742)(3600.00000,87.87931)(3630.00000,86.29084)(3660.00000,85.94520)(3690.00000,85.19381)(3720.00000,84.80751)(3750.00000,88.01857)(3780.00000,85.79981)(3810.00000,83.76155)(3840.00000,83.86903)(3870.00000,83.37543)(3900.00000,81.83492)(3930.00000,78.90784)(3960.00000,74.92460)(3990.00000,74.29432)(4020.00000,71.83504)(4050.00000,70.75226)(4080.00000,70.87882)(4110.00000,71.58584)(4140.00000,72.35453)(4170.00000,73.22304)(4200.00000,75.08178)(4230.00000,76.43616)(4260.00000,75.83276)(4290.00000,75.07109)(4320.00000,74.61528)(4350.00000,74.69135)(4380.00000,74.86615)(4410.00000,73.95190)(4440.00000,75.91949)(4470.00000,77.74198)(4500.00000,79.66643)(4530.00000,79.80302)(4560.00000,78.86226)(4590.00000,80.86121)(4620.00000,80.47312)(4650.00000,80.32865)(4680.00000,82.31187)(4710.00000,82.32191)(4740.00000,81.13287)(4770.00000,81.65140)(4800.00000,82.70398)(4830.00000,81.12702)(4860.00000,80.40946)(4890.00000,81.35145)(4920.00000,82.30211)(4950.00000,86.49550)(4980.00000,88.77785)(5010.00000,90.04188)};

            \addlegendentry{500 (198MiB)}
            
            \addplot[
                name path=1000,
                color=COL1000,
                style={ultra thick},
                solid,
                mark options={solid},
                ] coordinates
           {(1.00000,2373.61133)(30.00000,404.99806)(60.00000,234.63292)(90.00000,177.61332)(120.00000,146.74088)(150.00000,136.18836)(180.00000,138.57929)(210.00000,124.19511)(240.00000,139.48573)(270.00000,121.70090)(300.00000,124.80592)(330.00000,108.18059)(360.00000,107.21273)(390.00000,109.81193)(420.00000,96.87059)(450.00000,95.09996)(480.00000,92.07788)(510.00000,94.78796)(540.00000,91.42304)(570.00000,90.59099)(600.00000,99.55437)(630.00000,103.21678)(660.00000,103.26011)(690.00000,96.54145)(720.00000,89.40584)(750.00000,95.58014)(780.00000,95.77080)(810.00000,95.22734)(840.00000,93.86496)(870.00000,87.18125)(900.00000,85.38447)(930.00000,80.25709)(960.00000,79.33051)(990.00000,77.90560)(1020.00000,80.25768)(1050.00000,79.89504)(1080.00000,87.06692)(1110.00000,86.76932)(1140.00000,83.78626)(1170.00000,81.24526)(1200.00000,77.31975)(1230.00000,78.57182)(1260.00000,83.79856)(1290.00000,85.08432)(1320.00000,81.56777)(1350.00000,80.75487)(1380.00000,80.50867)(1410.00000,81.63377)(1440.00000,83.15584)(1470.00000,84.44799)(1500.00000,83.62342)(1530.00000,78.85648)(1560.00000,75.97231)(1590.00000,74.53401)(1620.00000,75.86561)(1650.00000,77.83939)(1680.00000,76.19597)(1710.00000,73.71978)(1740.00000,73.55442)(1770.00000,72.67833)(1800.00000,68.90103)(1830.00000,66.42651)(1860.00000,65.66209)(1890.00000,63.43815)(1920.00000,62.62412)(1950.00000,68.82730)(1980.00000,72.93467)(2010.00000,75.00690)(2040.00000,76.08498)(2070.00000,77.80063)(2100.00000,74.64908)(2130.00000,76.46753)(2160.00000,77.74116)(2190.00000,74.42423)(2220.00000,74.27477)(2250.00000,74.86312)(2280.00000,74.26490)(2310.00000,73.26394)(2340.00000,73.15415)(2370.00000,74.45243)(2400.00000,76.47255)(2430.00000,78.24368)(2460.00000,78.50371)(2490.00000,78.64772)(2520.00000,76.75092)(2550.00000,75.67867)(2580.00000,77.11592)(2610.00000,77.84324)(2640.00000,76.97045)(2670.00000,75.75366)(2700.00000,75.54408)(2730.00000,72.72992)(2760.00000,72.84960)(2790.00000,72.81066)(2820.00000,71.84825)(2850.00000,73.37313)(2880.00000,72.96505)(2910.00000,75.72641)(2940.00000,75.70758)(2970.00000,75.12956)(3000.00000,76.44350)(3030.00000,75.54375)(3060.00000,75.92097)(3090.00000,72.91989)(3120.00000,67.39742)(3150.00000,64.63973)(3180.00000,65.01229)(3210.00000,66.06020)(3240.00000,68.53675)(3270.00000,72.14607)(3300.00000,75.02437)(3330.00000,76.21880)(3360.00000,77.44915)(3390.00000,78.14457)(3420.00000,78.45280)(3450.00000,76.63466)(3480.00000,75.61740)(3510.00000,74.23435)(3540.00000,73.56241)(3570.00000,71.95511)(3600.00000,72.15242)(3630.00000,73.29907)(3660.00000,74.53944)(3690.00000,77.00529)(3720.00000,78.26966)(3750.00000,77.83637)(3780.00000,75.70284)(3810.00000,76.16715)(3840.00000,76.07839)(3870.00000,75.89663)(3900.00000,76.12423)(3930.00000,76.77432)(3960.00000,76.75938)(3990.00000,77.40265)(4020.00000,75.78893)(4050.00000,74.80091)(4080.00000,75.50619)(4110.00000,73.64360)(4140.00000,73.23788)(4170.00000,71.96050)(4200.00000,71.46481)(4230.00000,70.71920)(4260.00000,70.31542)(4290.00000,69.08039)(4320.00000,70.86065)(4350.00000,68.99377)(4380.00000,68.68786)(4410.00000,69.36053)(4440.00000,70.81020)(4470.00000,71.61988)(4500.00000,72.20527)(4530.00000,71.77245)(4560.00000,70.86666)(4590.00000,69.92979)(4620.00000,71.63699)(4650.00000,71.75230)(4680.00000,71.22259)(4710.00000,70.91421)(4740.00000,71.17510)(4770.00000,69.78251)(4800.00000,69.52046)(4830.00000,70.75894)(4860.00000,70.37384)(4890.00000,68.63107)(4920.00000,68.36112)(4950.00000,68.54410)(4980.00000,67.51878)(5010.00000,68.74315)};

            \addlegendentry{1000 (396MiB)}

            \addplot[
                name path=All,
                color=COLsdcfr,
                style={ultra thick},
                solid,
                mark options={solid},
                ] coordinates
           {(1.00000,2373.61133)(30.00000,381.45848)(60.00000,226.78933)(90.00000,186.00441)(120.00000,154.39717)(150.00000,141.66960)(180.00000,141.68800)(210.00000,138.73936)(240.00000,126.47322)(270.00000,116.73802)(300.00000,111.71467)(330.00000,110.83163)(360.00000,116.82523)(390.00000,117.34464)(420.00000,110.26822)(450.00000,111.19175)(480.00000,102.92437)(510.00000,94.61301)(540.00000,94.57735)(570.00000,93.64876)(600.00000,92.58759)(630.00000,97.06905)(660.00000,102.77252)(690.00000,104.18112)(720.00000,98.43136)(750.00000,99.75137)(780.00000,95.91319)(810.00000,91.96688)(840.00000,89.13601)(870.00000,91.49228)(900.00000,89.94055)(930.00000,86.22172)(960.00000,92.25098)(990.00000,88.66829)(1020.00000,88.72790)(1050.00000,89.12872)(1080.00000,88.28701)(1110.00000,86.69406)(1140.00000,87.72501)(1170.00000,86.91777)(1200.00000,87.29684)(1230.00000,86.57532)(1260.00000,88.57685)(1290.00000,88.47544)(1320.00000,86.70290)(1350.00000,87.87214)(1380.00000,83.61359)(1410.00000,84.16934)(1440.00000,82.24647)(1470.00000,78.68258)(1500.00000,74.60876)(1530.00000,74.69924)(1560.00000,76.55276)(1590.00000,79.35901)(1620.00000,78.04755)(1650.00000,73.96620)(1680.00000,74.43080)(1710.00000,76.58732)(1740.00000,80.31264)(1770.00000,79.68849)(1800.00000,79.13798)(1830.00000,77.16058)(1860.00000,74.77252)(1890.00000,72.96379)(1920.00000,71.72141)(1950.00000,69.74870)(1980.00000,69.87396)(2010.00000,69.30000)(2040.00000,68.46460)(2070.00000,67.87156)(2100.00000,68.05638)(2130.00000,66.97242)(2160.00000,67.70881)(2190.00000,70.32812)(2220.00000,71.32339)(2250.00000,69.84151)(2280.00000,67.86035)(2310.00000,66.97885)(2340.00000,67.61827)(2370.00000,68.50820)(2400.00000,68.02876)(2430.00000,67.91904)(2460.00000,67.75500)(2490.00000,66.82547)(2520.00000,65.76090)(2550.00000,64.77975)(2580.00000,65.79037)(2610.00000,66.23838)(2640.00000,65.03359)(2670.00000,67.64484)(2700.00000,66.96325)(2730.00000,66.37604)(2760.00000,66.63423)(2790.00000,66.62085)(2820.00000,67.59369)(2850.00000,68.70652)(2880.00000,68.26175)(2910.00000,67.74465)(2940.00000,68.05917)(2970.00000,67.21598)(3000.00000,67.27689)(3030.00000,66.70341)(3060.00000,65.48807)(3090.00000,65.87108)(3120.00000,65.48552)(3150.00000,63.23693)(3180.00000,65.03636)(3210.00000,64.85020)(3240.00000,63.72582)(3270.00000,62.89599)(3300.00000,62.11175)(3330.00000,61.56929)(3360.00000,61.08648)(3390.00000,61.25096)(3420.00000,61.65993)(3450.00000,62.00178)(3480.00000,61.78718)(3510.00000,61.38363)(3540.00000,60.40469)(3570.00000,59.38618)(3600.00000,58.32509)(3630.00000,58.71773)(3660.00000,59.46145)(3690.00000,59.93789)(3720.00000,60.57844)(3750.00000,60.66835)(3780.00000,59.72911)(3810.00000,59.94697)(3840.00000,60.55234)(3870.00000,60.37849)(3900.00000,60.20205)(3930.00000,59.57906)(3960.00000,58.71832)(3990.00000,58.68466)(4020.00000,58.67017)(4050.00000,59.18573)(4080.00000,60.37195)(4110.00000,60.83474)(4140.00000,61.12495)(4170.00000,61.44301)(4200.00000,61.19049)(4230.00000,60.16495)(4260.00000,59.78807)(4290.00000,59.99441)(4320.00000,59.26126)(4350.00000,58.99326)(4380.00000,59.14012)(4410.00000,58.23881)(4440.00000,57.67924)(4470.00000,57.02022)(4500.00000,57.63453)(4530.00000,57.81806)(4560.00000,58.09892)(4590.00000,58.21695)(4620.00000,57.47830)(4650.00000,57.88988)(4680.00000,57.76441)(4710.00000,57.96801)(4740.00000,58.38671)(4770.00000,58.90518)(4800.00000,59.51113)(4830.00000,59.86625)(4860.00000,59.79977)(4890.00000,59.53355)(4920.00000,58.91913)(4950.00000,58.91983)(4980.00000,59.02260)};
            \addlegendentry{Infinite}

            \addplot[mark=none, COLZeroLine, samples=2222, forget plot] coordinates {
            (0,  0)
            (10000,  0)};

            \end{loglogaxis}
        \end{tikzpicture}
}
\caption{Empirical analysis of SD-CFR in Leduc Hold'em Poker. Results in Figure \ref{fig:explLeduc} and \ref{fig:explResSampl} are averaged over five and three runs, respectively.}
\end{center}
\end{figure}
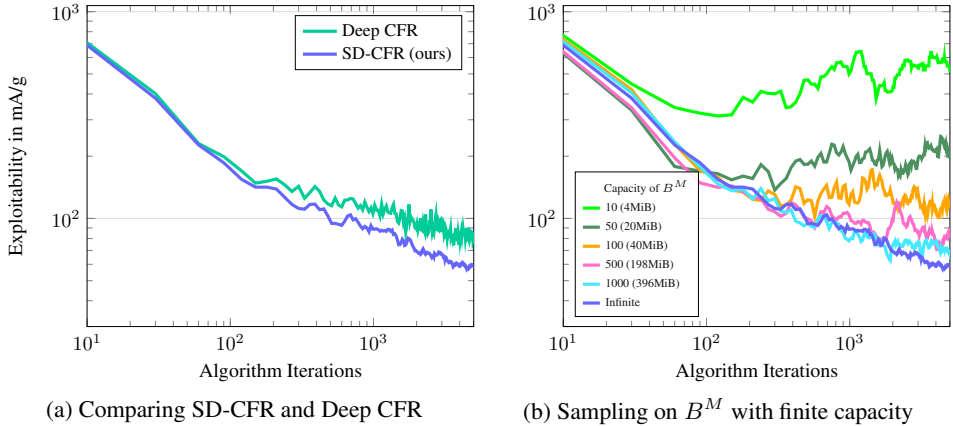

Figure \ref{fig:explLeduc} shows the exploitability (i.e. loss against a worst-case opponent) of SD-CFR and Deep CFR in Leduc Poker ~\citep{Leduc} measured in \textit{milli-antes per game (mA/g)}.

In Leduc Poker, players start with an infinite number of chips. The deck has six cards of two suits $\{a, b\}$ and three ranks $\{J, Q, K\}$. There are two betting rounds: preflop and flop. After the preflop, a card is publicly revealed. At the start of the game, each player adds 1 chip, called the ante, to the pot and is dealt one private card. There are at most two bets/raises per round, where the bet-size is fixed at 2 chips in the preflop, and 4 chips after the flop is revealed. If no player folded, the winner is determined via hand strength. If a player made a pair with the public card, they win. Otherwise $K > Q > J$. If both players hold a card of the same rank, the pot is split.

Hyperparameters are chosen to favour Deep CFR as the neural networks and buffers are very large in relation to the size of the game. Yet, we find that SD-CFR minimizes exploitability better than Deep CFR. Exact hyperparameters can be found in the supplementary material.

Although we concluded that storing all value networks is feasible, we analyze the effect of reservoir sampling on $B^M$ in Figure \ref{fig:explResSampl} and find it leads to plateauing and oscillation, at least up to $|B^M| = 1000$.

\subsection{One-on-One in 5-Flop Hold'em Poker (5-FHP) against Deep CFR}
\begin{wrapfigure}{R}{0.5\textwidth}
    \begin{center}
        \begin{tikzpicture}[scale=0.8]
        \pgfplotsset{
            legend style={font=\footnotesize}
            }
            \begin{axis}[
                xlabel={Algorithm Iterations},
                ylabel={SD-CFR's winnings in mbb/g},
                xmin=20, xmax=310,
                ymin=-10, ymax=47,
                xtick={30, 60, 90, 120, 150, 180, 210, 240, 270, 300},
                ytick={-30, -20, -10, 0, 10, 20, 30, 40, 50, 60},
                legend pos=north east,
                ymajorgrids=true,
                grid style={thin, solid, COLgrid},
            ]
            
            \addplot[
                name path=run1M,
                color=COLR1,
                dashed,
                style={thick},
                mark=*,
                mark options={solid},
                ] coordinates 
            {(30.00000,41.18000)(60.00000,15.81000)(90.00000,19.08000)(120.00000,7.49100)(150.00000,13.17000)(180.00000,7.62200)(210.00000,-1.14600)(240.00000,13.30000)(270.00000,12.10000)(300.00000,11.67000)};
            \addlegendentry{Run 1}

            \addplot[
                name path=run2M, 
                color=COLR2,
                dashed,
                style={thick},
                mark=*,
                mark options={solid},
                ] coordinates 
            {(30.00000,41.16000)(60.00000,9.69100)(90.00000,2.23900)(120.00000,7.21400)(150.00000,-1.31300)(180.00000,5.41000)(210.00000,10.17000)(240.00000,9.13800)(270.00000,9.66500)(300.00000,8.32700)};
            \addlegendentry{Run 2}

            \addplot[
                name path=run3M,
                color=COLR3,
                dashed,
                style={thick},
                mark=*,
                mark options={solid},
                ] coordinates 
            {(30.00000,28.08000)(60.00000,10.55000)(90.00000,10.12000)(120.00000,7.50100)(150.00000,10.12000)(180.00000,1.98400)(210.00000,11.71000)(240.00000,4.90100)(270.00000,8.28900)(300.00000,9.15500)};
            \addlegendentry{Run 3}

            \addplot[
                name path=run4M,
                color=COLR4,
                dashed,
                style={thick},
                mark=*,
                mark options={solid},
                ] coordinates 
            {(30.00000,22.50000)(60.00000,5.45000)(90.00000,7.07100)(120.00000,10.82000)(150.00000,10.99000)(180.00000,11.44000)(210.00000,6.02000)(240.00000,6.66000)(270.00000,-1.14000)(300.00000,9.99000)};
            \addlegendentry{Run 4}

            \addplot[
                name path=run4M,
                color=COLR5,
                dashed,
                style={thick},
                mark=*,
                mark options={solid},
                ] coordinates 
            {(30.00000,18.28000)(60.00000,9.21000)(90.00000,2.15000)(120.00000,0.44000)(150.00000,1.11000)(180.00000,16.35000)(210.00000,8.40000)(240.00000,11.58000)(270.00000,3.71000)(300.00000,2.88000)};
            \addlegendentry{Run 5}

            \addplot[
                name path=Mean,
                color=COLavg, 
                style={ultra thick},
                solid,
                mark options={solid},
                smooth
                ] coordinates 
            {(30.00000,30.24000)(60.00000,10.14220)(90.00000,8.13200)(120.00000,6.69320)(150.00000,6.81540)(180.00000,8.56120)(210.00000,7.03080)(240.00000,9.11580)(270.00000,6.52480)(300.00000,8.40440)};
            \addlegendentry{Mean}

            \addplot[mark=none, COLZeroLine, samples=2222, forget plot] coordinates {
            (0,  0)
            (10000,  0)
            };
            
            \end{axis}
        \end{tikzpicture}
    \caption{\textbf{One-on-One performance} of Single Deep CFR vs. Deep CFR. Dashed lines are independent algorithm runs. All evaluations have 95$\%$ confidence intervals between $\pm 5.4$ and $\pm 6.51$ and are the average result of 3M poker hands each.}
    \label{fig:oooresults}
    \end{center}
\end{wrapfigure}
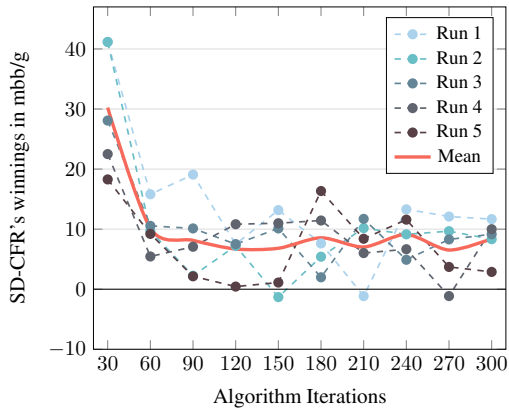

Figure \ref{fig:oooresults} shows the results of one-one-one matches between SD-CFR and Deep CFR in 5-Flop Hold'em Poker (5-FHP). 5-FHP is a large poker game similar to regular FHP ~\citep{brown2018deep}, which was used to evaluate Deep CFR ~\citep{brown2018deep}. The only difference is that 5-FHP uses five instead of three flop cards, forcing agents to abstract and generalize more. For details on FHP, please refer to ~\citep{brown2018deep}. The neural architecture is as ~\cite{brown2018deep}. Both algorithms again \textit{share the same value networks} during each training run. Like ~\cite{brown2018deep}, $B^v$ and $B^s$ have a capacity of 40 million per player. On each iteration, we run a batch of 300,000 external sampling traversals and train a value network from scratch using a batch size of 10,240 for 4,000 updates. Average strategy networks are trained with a batch size of 20,480 for 20,000 updates. SD-CFR's $B^M$ stores all value networks, requiring 120MB of disk space, while each $B^s$ needs around 25GB of memory during training.

The y-axis plots SD-CFR's average winnings against Deep CFR in \textit{milli-big blinds per game (mbb/g)} measured every 30 iterations. For reference, 10 mbb/g is considered a good margin between humans in Heads-Up Limit Hold'em (HULH), a game with longer action sequences, but similar minimum and maximum winnings per game as 5-FHP. Measuring the performance on iteration $t$ compares how well the SD-CFR averaging procedure would do against the one of Deep CFR if the algorithm stopped training after $t$ iterations

$B^s$ reached its maximum capacity of 40 million for both players by iteration 120 in all runs. Before this point, SD-CFR defeats Deep CFR by a sizable margin, but even after that, SD-CFR clearly defeats Deep CFR.

\subsubsection{Comparing strategies}
\begin{table}
    \vskip 0.15in
    \begin{center}
        \begin{small}
            \begin{sc}
                \begin{tabular}{lcccr}
                \toprule
                Depth   & Round     & Dif Mean            & Dif STD  & N         \\
                \midrule
                0       & PF        & 0.012$\pm$ 0.0001   & 0.017    & 200k      \\
                1       & PF        & 0.013$\pm$ 0.0001   & 0.018    & 100k      \\
                2       & FL        & 0.052$\pm$ 0.0003   & 0.048    & 80k       \\
                3       & FL        & 0.083$\pm$ 0.0005   & 0.075    & 83k       \\
                4       & FL        & 0.113$\pm$ 0.0011   & 0.109    & 37k       \\
                5       & FL        & 0.175$\pm$ 0.0057   & 0.206    & 5k        \\
                \bottomrule
                \end{tabular}
        \end{sc}
        \end{small}
    \end{center}
    \caption{\textbf{Disagreement between SD-CFR's and Deep CFR's average strategies}. "DEPTH": number of player actions up until the measurement, "ROUND": PF=Preflop, FL=Flop, "DIF MEAN": mean and 95\% confidence interval of the absolute differences between the strategies over the "N" occurrences. "DIF STD": approximate standard deviation of agreement across information sets.}
    \label{tab:stratComp}
\end{table}
We analyze how far the average strategies of SD-CFR and Deep CFR are apart at different depths of the tree of 5-FHP. In particular, we measure
\begin{equation*}
    \frac{1}{2}\sum_{i \in {1, 2}}
        (\E_{I_i \sim \bar{\sigma}^T_i}
            \sum_{a \in A(I)}
                |{\bar{\sigma}^{T, SD}_i(I, a) - \bar{\sigma}^{T, \hat{S}}_i(I, a)} |)
\end{equation*}
We ran 200,000 trajectory rollouts for each player, where player $i$ plays according to SD-CFR's average strategy $\bar{\sigma}^{T, SD}_i$ and $-i$ plays uniformly random. Hence, we only evaluate on trajectories on which the agent should feel comfortable. The two agents again share the same value networks and thus approximate the same equilibrium. We trained for 180 iterations, a little more than it takes for $B^s$ and $B^v$ to be full for both players. Table \ref{tab:stratComp} shows that Deep CFR's approximation is good on early levels of the tree but has a larger error in information sets reached only after multiple decision points.

\section{Related Work}
\textit{Regression CFR (R-CFR)} ~\citep{RegCFR} applies regression trees to estimate regret values in CFR and CFR$^+$. Unfortunately, despite promising expectations, recent work failed to apply R-CFR in combination with sampling ~\citep{srinivasan2018actor}. \textit{Advantage Regret Minimization (ARM)} ~\citep{ARM} is similar to R-CFR but was only applied to single-player environments. Nevertheless, ARM did show that regret-based methods can be of interest in imperfect information games much bigger, less structured, and more chaotic than poker.

\textit{DeepStack} ~\citep{deepstack} was the first algorithm to defeat professional poker players in one-on-one gameplay of Heads-Up No-Limit Hold'em Poker (HUNL) requiring just a single GPU and CPU for real-time play. It accomplished this through combining real-time solving with counterfactual value approximation with deep networks. Unfortunately, DeepStack relies on tabular CFR methods without card abstraction to generate data for its counterfactual value networks, which could make applications to domains with many more private information states than HUNL has difficult.

\textit{Neural Fictitious Self-Play (NFSP)} ~\citep{NFSP} was the first algorithm to soundly apply deep reinforcement learning from single trajectory samples to large extensive-form games. While not showing record-breaking results in terms of exploitability, NFSP was able to learn a competitive strategy in Limit Texas Hold'em Poker over just 14 GPU/days. Recent literature elaborates on the convergence properties of multi-agent deep reinforcement learning ~\citep{unified} and introduces novel actor-critic algorithms ~\citep{srinivasan2018actor} that have similar convergence properties as NFSP and SD-CFR.

\section{Future Work}
So far, Deep CFR was only evaluated in games with three player actions. Since external sampling would likely be intractable in games with tens or more actions, one could employ outcome sampling ~\citep{MCCFR}, robust sampling ~\citep{doubleDCFR}, Targeted CFR ~\citep{jackson2017targeted}, or average-strategy-sampling ~\citep{AvrgStratSamplMC} in such settings.

To avoid action translation after training in an action-abstracted game, continuous approximations of large discrete action-spaces where actions are closely related (e.g. bet-size selection in No-Limit Poker games, auctions, settlements, etc.) could be of interest. This might be achieved by having the value networks predict parameters to a continuous function whose integral can be evaluated efficiently. The iteration-strategy could be derived by normalizing the advantage clipped below 0. The probability of action $a$ could be calculated as the integral of the strategy on the interval corresponding to $a$ in the discrete space.

Given a few modifications to its neural architecture and sampling procedure, SD-CFR could potentially be applied to much less structured domains than poker such as those that deep reinforcement learning methods like PPO ~\citep{PPO} are usually applied to. A first step on this line of research could be to evaluate whether SD-CFR is preferred over approaches such as ~\citep{srinivasan2018actor} in these settings.

\section{Conclusions}
We introduced \textit{Single Deep CFR (SD-CFR)}, a new variant of CFR that uses function approximation and partial tree traversals to generalize over the game's state space. In contrast to previous work, SD-CFR extracts the average strategy directly from a buffer of value networks from past iterations. We show that SD-CFR is more attractive in theory and performs much better in practise than Deep CFR.

\subsubsection*{Acknowledgments}
Eric thanks Johannes Heinrich for his mentorship and for valuable discussions throughout the project of which this paper was a part. Furthermore, we appreciate Alexander Mandt's work on our distributed computing set-up, and thank Sebastian De Ro for his contribution to auxiliary tools in our codebase. Michael Johanson pointed out that the discussed averaging method generalizes to tabular CFR and provided extensive feedback. HTBLVA Spengergasse supported this project in an early stage by allowing access to their infrastructure. Lastly, we thank Yannis Wells, Lora Naydenova, and James Read for proof-reading and helpful suggestions.

\section*{Code}
\url{https://github.com/EricSteinberger/Deep-CFR} implements Deep CFR and SD-CFR, and provides scripts to reproduce the results presented in this paper.

\newpage
\appendix
\section{Hyperparameters of experiments performed in Leduc Hold'em Poker}
$B^v$ and $B^s$ have a capacity of 1 million for each player. On each iteration, data is collected over 1,500 external sampling traversals and a new value network is trained to convergence (750 updates of batch size 2048), initialized randomly at $t<2$ and with the weights of the value net from iteration $t-2$ afterwards. Average-strategy networks are trained to convergence (5000 updates of batch size 2048) always from a random initialization. All networks used for this evaluation have 3 fully-connected layers of 64 units each, which adds up to more parameters than Leduc Hold'em has states. All other hyperparameters were chosen as ~\cite{brown2018deep}.

\section{Rules of Leduc Hold'em Poker}
Leduc Hold'em Poker is a two-player game, were players alternate seats after each round. At the start of the game, both players add 1 chip, the \textit{ante}, to the pot and are dealt a \textit{private card} (unknown to the opponent) from a deck consisting of 6 cards: \{A, A, B, B, C, C\}. There are two rounds: \textit{pre-flop} and \textit{flop}. The game starts at the pre-flop and transitions to the flop after both players have acted and wagered the same number of chips. At each decision point, players can choose an action from a subset of \{\textit{fold},\textit{call}, \textit{raise}\}. When a player folds, the game ends and all chips in the pot are awarded to the opponent. Calling means matching the opponent's raise. The first player to act in a round has the option of \textit{checking}, which is essentially a call of zero chips. Their opponent can then bet or also check. When a player raises, he adds more chips to the pot than his opponent wagered so far. In Leduc Hold'em, the number of raises per round is capped at 2. Each raise adds 2 additional chips in the pre-flop round and 4 in the flop round. On the transition from pre-flop to flop, one card from the remaining deck is revealed publicly. If no player folded and the game ends with a player calling, they show their hands and determine the winner by the rule that if a player's private card matches the flop card, they win. Otherwise the player with the higher card according to $A\>B\>C$ wins.

\section{Deep CFR performs well on early iterations in some games}
We conducted experiments searching to investigate the harm caused by the function approximation of $\hat{S}$. We found that in variants of Leduc Hold'em ~\citep{Leduc} with more that 3 ranks and multiple bets, the performance between Deep CFR and SD-CFR was closer. Below we plot the exploitability curves of the early iterations in a variant of Leduc that uses a deck of 12 ranks and allows a maximum of 6 instead of 2 bets per round.

We believe the smaller difference in performance is due to the equilibrium in this game being less sensitive to small differences in action probabilities, while the game is still small enough to see every state often during training. In vanilla Leduc, slight deviations from optimal play give away a lot about one's private information as there are just three distinguishable cards. In contrast, this variant of Leduc, despite having more states, might be less susceptible to approximation error as it has 12 distinguishable cards but similarly simple rules.

For the plot below, we ran Deep CFR and SD-CFR with shared value networks, where all buffers have a capacity of 4 million. On each iteration, data is collected over 8,800 external sampling traversals and a new value network is trained to convergence (1200 updates of batch size 2816), initialized randomly at $t<2$ and with the weights of the value net from iteration $t-2$ afterwards. Average-strategy networks are trained to convergence (10000 updates of batch size 5632) from a random initialization. The network architecture used is as ~\cite{brown2018deep}, differing only by the card-branch having 64 units per layer instead of 192.

\begin{figure}[!htb]
    \centering
        \begin{tikzpicture}[scale=0.97]
        \pgfplotsset{
            legend style={font=\footnotesize}
            }
            \begin{loglogaxis}[
                legend cell align={left},
                xlabel={Algorithm Iterations},
                ylabel={Exploitability in mA/g},
                xmin=10, xmax=630,
                ymin=80, ymax=1100,
                legend pos=north east,
                legend cell align={left},
                ymajorgrids=true,
                grid style={thin, solid, COLgrid},
            ]
            
            \addplot[
                name path=DeepCFR,
                color=COLdeepcfr, 
                style={ultra thick},
                solid,
                mark options={solid},
                ] coordinates 
            {(1.00000,4013.10112)(30.00000,435.70656)(60.00000,283.06968)(90.00000,215.37766)(120.00000,194.69075)(150.00000,182.82509)(180.00000,159.86010)(210.00000,150.11700)(240.00000,159.70840)(270.00000,146.25542)(300.00000,144.90193)(330.00000,145.09727)(360.00000,142.48978)(390.00000,144.87162)(420.00000,131.89809)(450.00000,141.53180)(480.00000,142.51350)(510.00000,135.05717)(540.00000,134.62591)(570.00000,140.17260)(600.00000,137.19639)(630.00000,127.95697)};
            \addlegendentry{Deep CFR}
            
            \addplot[
                name path=SDCFR,
                color=COLsdcfr,
                style={ultra thick},
                solid,
                mark options={solid},
                smooth
                ] coordinates
           {(1.00000,4157.66504)(30.00000,448.18769)(60.00000,256.64786)(90.00000,210.48993)(120.00000,186.10636)(150.00000,176.38943)(180.00000,160.22078)(210.00000,153.34598)(240.00000,153.06370)(270.00000,145.60730)(300.00000,138.67237)(330.00000,139.61758)(360.00000,136.08221)(390.00000,135.26280)(420.00000,128.87237)(450.00000,129.73153)(480.00000,131.63798)(510.00000,134.30363)(540.00000,131.80290)(570.00000,132.60036)(600.00000,128.46723)(630.00000,121.68144)};
            \addlegendentry{SD-CFR (ours)}
        
            \addplot[mark=none, COLZeroLine, samples=2222, forget plot] coordinates {
            (0,  0)
            (10000,  0)
            };
            
            \end{loglogaxis}
        \end{tikzpicture}
    \caption{\textbf{Exploitability} of Single Deep CFR and Deep CFR averaged over five runs.}
    \label{fig:explBigLeduc}
\end{figure}
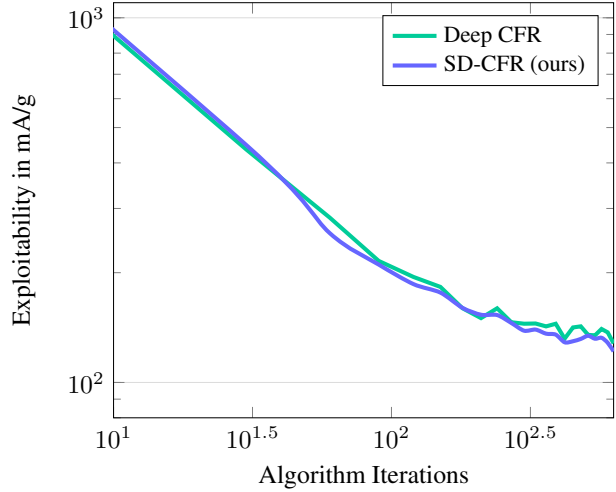

\section{Proof of Theorem 1}
    \begin{proof}
        Let $I$ be any information set in $\mathcal{I}_i$. Assuming that $0 < \pi_i^{\sigma^t}(I) < 1$. Recall that external sampling samples only one action for player $i$ and $chance$ at any decision point, when $-i$ is the traverser. Since $(1 - \pi_i^{\sigma^t}(I))^K > 0$ for any finite number of traversals $K$ per iteration, we cannot guarantee that $I$ will be visited. If $I$ is not visited despite $\pi_i^{\sigma^t}(I) > 0$, the contribution of $\sigma^t_i$ to $\bar{\sigma}^T_i(I)$ is not represented in $B_i^s$.
        
        For the second argument, we assume that $K=\infty$. Let $I$ again be any information set in $\mathcal{I}_i$ in which $|A(I)| > 1$. Assume that $\pi_i^{\sigma^t}(I)$ is irrational and that $\pi_i^{\sigma^j}(I)$ is rational. Clearly, because its capacity is finite, $B_i^s$ could not reflect the ratio between $\pi_i^{\sigma^t}(I)$ and $\pi_i^{\sigma^j}(I)$ correctly through the frequency of the appearance of samples from iterations $t$ and $j$, regardless of the number of traversals. Furthermore, in games where the number of members in the set \begin{equation*}
        \{I \in \mathcal{I}_i : |A(I)| > 1, \pi_i^{\sigma^t}(I) > 0\}
        \end{equation*}
        is bigger than the capacity of $B^s_i$, not every $I \in \Tilde{I}$ can fit into $B_i^s$ on iteration $t$, also making $B_i^s$ an incomplete representation of $\bar{\sigma}^T_i(I)$.
    \end{proof}

\section{Proof of Theorem 2}
    \begin{proof}
    Let $B_i^M$ be a buffer of all value networks up to iteration $T$ belonging to player $i$.
    
    Since $\hat{D}_i^t(I, a) = D^t_i(I, a)$ for all $I \in \mathcal{I}_i$ and all $a \in A(I)$ by assumption, 
    \begin{equation} \label{eq:PROOF_DCFRimmStrat}
        \sigma^{t+1}_i(I, a) =
        \begin{cases}
            \frac{D_i^t(I, a)_{+}}{\sum_{\Tilde{a} \in A(I)} D_i^t(I, \Tilde{a})_{+}}  & \text{if} \sum_{\Tilde{a}} D_i^t(I, \Tilde{a})_{+} > 0 \\
            \frac{1}{|A(I)|}           & \text{otherwise} \\
        \end{cases}
    \end{equation}
    can be restated in terms of $\hat{D}_i^t(I, a)$.
    
    By definition, 
    \begin{equation} \label{eq:PROOF_Reach_ii} 
        \pi_{i}^{\sigma^t}(I) = \prod_{I' \in I, P(I')=i, a': I'\to I} \sigma^{t}_i(I', a')
    \end{equation}
    Since all $\sigma^{t}_i$ have no error by assumption, SD-CFR's recomputation of $\pi_{i}^{\sigma^t}(I)$ and hence also $\bar{\sigma}^T_i(I, a)$ are exact for any $I \in \mathcal{I}_i$ and all $a \in A(I)$.
    
    To show this for trajectory-sampling SD-CFR, consider a trajectory starting at the tree's root $\phi$ leading to an information set in $I \in \mathcal{I}_i$. Since $\sigma^t_i$ can be deduced from $\hat{D}^t_i$ as before, $B_i^M$ can be seen as a buffer of iteration-strategies. Let $f\colon I\to a$ be a function that first chooses a $\sigma^t_i \in B_i^M$, where each $\sigma^t_i$ is assigned a sampling weight of $t\pi_i^{\sigma}(I)$. $f$ then returns an action sampled from the distribution $\sigma^t_i(I)$. Since $f$ weights strategies like the numerator of the definition
    \begin{equation} \label{eq:PROOF_DCFRavgStrat}
        \bar{\sigma}^T_i(I, a) = \frac{\sum_{t=1}^{T} (t \pi_{i}^{\sigma^t}(I)  \sigma_{i}^{t}(I, a))} {\sum_{t=1}^{T}     (t\pi_{i}^{\sigma^t}(I))}
    \end{equation}
    executing $f(I)$ is equivalent to sampling directly from $\bar{\sigma}^T_i$.
    
    Note that $\pi_i^{\sigma}(\phi) = 1$ for all $\sigma$. Thus, $f(\phi)$ would choose a given $\sigma_i^t \in B_i^M$ with sampling weight $t$. This is what trajectory-sampling SD-CFR does at $\phi$. For each information set $I'$ from $\phi$ until the end of the trajectory, SD-CFR plays using the same iteration-strategy selected at $\phi$. Thus, SD-CFR will reach each information set $I$ with a probability proportional to $\pi_i^{\sigma^t}(I)$ conditional on knowing which iteration-strategy was selected. Combining these facts, we see that the assigned weight of $\sigma^t_i$ in any $I$ is $t\pi_i^{\sigma^t}(I)$ for any $t$ up to $T$. It follows that the probability of $\sigma^t_i$ being the acting policy in any $I$ is
    \begin{equation*}
        \frac{t\pi_i^{\sigma^t}(I)} {\sum^T_{t'=1} (t'\pi_i^{\sigma^{t'}}(I))}
    \end{equation*}
    Since this is equivalent to the weighting scheme between iteration-strategies in the definition of $\bar{\sigma}^T_i$, trajectory-sampling SD-CFR samples correctly from $\bar{\sigma}^T_i$.
    
    Moreover, because the opponent does not know which $\sigma^t_i$ is the acting policy, this result also shows that an opponent cannot tell whether the agent is using this sampling method or following an explicitly computed $\bar{\sigma}_i^T$
    \end{proof}
    
\end{document}